\documentclass[aps,pre,twocolumn,groupedaddress,showpacs]{revtex4-1}
\usepackage{graphicx}
\usepackage{subfig}
\usepackage{enumerate}
\usepackage{multirow}
\usepackage{amsmath,amsthm,amssymb}

\begin{document}


\title{Small-world structure induced by spike-timing-dependent plasticity in networks with critical dynamics}
    

	\author{Victor Hernandez-Urbina}
	\email[]{j.v.hernandez-urbina@ed.ac.uk}
	\author{J. Michael Herrmann}
	\affiliation{Institute of Perception, Action and Behaviour, University of Edinburgh. EH8 9AB \\
		United Kingdom.}	


	\date{\today}

	\begin{abstract}
			The small-world property in the context of complex networks implies structural benefits to the 
			processes taking place within a network, 
			such as optimal information transmission and robustness. 
			In this paper, we study a model network of integrate-and-fire neurons that are subject to 
			activity-dependent synaptic plasticity. We find the learning rule that gives rise to a 
			small-world structure when the collective dynamics of the system reaches a critical state 
			which is characterised by power-law distributions of activity clusters.
			Moreover, by analysing the motif profile of the networks, we observe that bidirectional
			connectivity is impaired by the effects of this type of plasticity.
	\end{abstract}


	\maketitle

	\section{Introduction}
		\label{intro}

		In their seminal paper, Watts and Strogatz~\cite{watts1998collective} described a class of 
		networks that lie halfway between completely random
		and regular networks.
		This class of networks is characterised by a small average path length (which is typical for dense
		random networks)
		and an average clustering coefficient significantly larger than expected by chance (as in many regular lattices). 
		However, the notion of a special class of small-world networks is somewhat misleading and it seems more appropriate to 
		speak of a \emph{small-world property}, i.e. this property is not a binary but a gradual one and is stronger in some networks
		as compared to others as can be expressed by quantitative 
		measures~\cite{humphries2008network}.
		
		The small-world property offers a structural benefit to the processes taking place within the 
		network, e.g.~with respect to information transmission 
		by speeding-up the communication among distant nodes 
		in a regular network that can be reached via short-cuts in the small-world case.
		Moreover, this class of networks is a suitable model
	    in social, technological 
		and biological contexts etc.~\cite{humphries2008network}.
		For example, it has been observed that the electrical power grid network of western United 
		States, the co-stardom network comprising movie actors and their collaborations and the neural 
		network of the nematode worm \emph{C.~elegans} have the structural properties 
		of low average path length along with a high average clustering 
		coefficient~\cite{watts1998collective}.

		In another seminal paper, Barab\'asi and Albert~\cite{barabasi1999emergence} proposed a 
		model, known as the \emph{BA model}, to explain the emergence of scale invariance in the degree distribution 
		e.g.~in the World Wide Web (WWW)~\cite{barabasi2002linked}. 
		In these \emph{scale-free networks} the 
		probability $P(k)$ that a node connects to $k$ other nodes follows a power-law 
		$P(k)\sim k^{-\gamma}$~\cite{barabasi1999emergence}. It implies the existence of many poorly 
		connected nodes coexisting with very few but 
		not negligible massively connected nodes \emph{hubs}. 
		Scale-invariant degree distribution and the small-world
		property are by no means exclusive and in fact many
		scale-invariant networks are also small-world. 
		
		A look at the degree distribution of a network and the estimation of some network statistics like local clustering coefficients
		and shortest path lengths provide an insightful initial description of network structure.
		A more detailed
		picture of the network structure can be obtained by considering 
		the relative frequency of particular configurations of small 
		blocks in comparison to random networks~\cite{milo2002network}. 
		In particular, $3$-node subgraphs, the \emph{motifs}, describe
		the relationship among nodes taken in threes.
		Song et al.~\cite{song2005highly} observed that the motif distribution of acute slices taken 
		from the visual cortex of rats exhibits an overrepresentation
		of some $3$-node patterns when compared to random networks.
		Moreover, networks from a similar context share similarities in their motif profiles, so that 
		these $3$-node subgraphs might even define broad
		classes, or super-families, of networks~\cite{milo2002network,milo2004superfamilies}. 
		Similar motif distributions could point to similar dynamical processes in the network, e.g.
		food networks represent the flow of energy from bottom
		to top of the food web resulting in a particular motif configuration, whereas brain networks 
		represent a flow of information without necessarily implying a
		particular direction of the flow, which results in a different motif 
		profile~\cite{milo2002network}.
		
		When it comes to the dynamics occurring in a system that comprises a large number of simple elements interacting in a network, 
		a large number of studies have been dedicated to the occurrence of power-law behaviour and its relationship
		with the concept of \emph{self-organised criticality} (SOC)~\cite{bak1988self}.
		The concept of SOC has been suggested to explain the dynamics of phenomena as diverse as plate 
		tectonics~\cite{gutenberg1956magnitude}, piles of granular
		matter~\cite{frette1996avalanche}, forest fires~\cite{bak1990forest}, neuronal 
		avalanches~\cite{eurich2002finite} (see below), 
		and mass extinctions~\cite{bak1997nature}, among several others.
		SOC implies the existence of a critical point that becomes an attractor in the 
		collective dynamics of a system which resides thus
		at the boundary between two different phases of the system (e.g.~`order' and 
		`chaos'). SOC can be characterised by power-law event-size distributions, 
		divergence of the correlation length, critical slowing-down of
		the decay of perturbation as well as the existence of avalanche-like causation of events 
		in the system ~\cite{bak1988self}.
		
	        In neural systems, an activated neuron can trigger a chain reaction which 
        		takes the form of a cascade of activity.
	        These neuronal avalanches have been observed in biological neurons~\cite{beggs2003neuronal} 
        		as well as in models (see Sect.~\ref{model}).
		It has been reported that the distribution of avalanche sizes and avalanche durations can be 
		approximated by power laws with resp.~exponents $\gamma = -3/2$ and $\delta=-2$ 
		in the thermodynamic limit in globally-coupled networks~\cite{eurich2002finite,levina2007dynamical}.
		Interestingly, the theoretical predictions are in good agreement with experiemental results 
		in real brain tissue in local field potentials of cultured slices of rat cortex~\cite{beggs2003neuronal},
		and in the superficial cortical layers of awake, resting 
		primates~\cite{petermann2009spontaneous}.

		Critical dynamics of brain networks have been studied thoroughly in artificial models, 
		and it has been found that the critical regime implies several computational benefits for the 
		system, namely: optimal information transmission and 
        maximum dynamic range~\cite{kinouchi2006optimal}, maximum information 
        storage~\cite{haldeman2005critical,uhlig2013critical}, 
        stability of information transmission~\cite{bertschinger2004real}, among others.
	
        An important area of research in network science is focused on studying the mechanisms by 
        which nodes connect and disconnect during the development of a network.
		Complex networks often possess feedback mechanisms by which 
		the node dynamics and their interactions affect the structure of the network, which in turn alters the behaviour of the nodes etc.
		It is thus useful to consider both the dynamics \emph{on} 
		networks as well as the dynamics \emph{of} networks~\cite{gross2008adaptive,gross2009adaptive}.	
		
		In brain networks, synaptic plasticity refers to the structural changes that neuronal networks 
		undergo through the strengthening or weakening of synaptic connections in response to the 
		in-going activity.
		In particular, \emph{spike-timing dependent plasticity} (STDP) captures the existing temporal 
		correlations between the spikes of pre- and post-synaptic units
		resulting in a temporally asymmetric learning rule that has been proposed as a
		mechanism for learning and memory in the brain~\cite{bi1998synaptic}.
		STDP is a local rule that emphasises the precise timing of each individual 
		spike, it also incorporates and extends the essential mechanism of 
		Hebbian learning by including the notion of long-term potentiation (LTP) and long-term 
		depression (LTD) of synapses based on the differences in activation times for pre- and 
		post-synaptic neurons~\cite{bi2001synaptic}.
		
		In this paper we study the effects of STDP mechanisms over systems comprising integrate-and-fire neurons poised at criticality
		and whose structures follow a complex topology.
		In the following sections we report how these two concepts combined, that of STDP and that of criticality, have an effect on
		structural properties of the network which are summarised in the concept of the small-world property.

	\section{Model}
	\subsection{The Eurich model}\label{Sect:model}
		\label{model}
		The model consists of $N$ non-leaky
		integrate-and-fire nodes and was formulated for all-to-all connectivity~\cite{eurich2002finite}, but here we will also consider heterogeneous directed networks.
		
		In the model, every node $j$ is characterised by a membrane potential $h_{j}$ which is a 
		continuous variable updated in discrete time according to 
		the following equation:
		\begin{equation}
			h_{j}(t+1) = h_{j}(t) + \sum_{i = 1}^{N} A_{ij} w_{ij}s_{i}(t) + I_{ext}
		\end{equation}
		 \noindent where $A$ represents the adjacency matrix with entries $A_{ij} = 1$ if node $i$ sends 
		 an edge to node $j$, and $A_{ij} = 0$ otherwise;
		 $w_{ij}$ denotes the synaptic strength from node $i$ to node $j$;  $s_{i}(t)\in \{1,0\}$ 
		 represents the state of node $i$ (active or quiescent, respectively) 
		 at time $t$; and $I_{ext}$ denotes external input which is supplied to a node depending on the 
		 state of the system at time $t$.
		 If there is no activity at time $t$, then a node is chosen uniformly at random and its membrane 
		 potential is increased by a fixed amount through the variable $I_{ext}$ that
         represents the external driving of the system.
		 When the potential exceeds a neural threshold, i.e.~$h_{i}(t)\geq\theta$, the node becomes active, $s_{i}(t) = 1$. 
		 Afterwards, this node is reset, i.e.~$h_{i} = 0$.
		 
        In the simulations we study networks of different sizes and with the following network topologies:
        \begin{enumerate}[\it i.]
            \item fully connected,
            \item random, 
            \item scale-free with low mean clustering coefficient (CC) and power-law out-degree 
            distribution, 
            \item scale-free with high CC and power-law out-degree distribution, 
            \item scale-free with low CC and power-law in-degree distribution,
            \item scale-free with high CC and power-law in-degree distribution.
        \end{enumerate}
        It is worth mentioning that for the case of random and scale-free structures, the networks 
        considered possess the same number of edges per system size, which results in the same average 
        connectivity per network. 
        In other words, as they have the same number of edges, their structure results from a 
        permutation of the edges.
		For a random network, edges are inserted independently as in the Erd\"os-Renyi model~\cite{newman2003structure},
		for scale-free networks we follow Ref.~\cite{holme2002growing}.
		The latter algorithm achieves tunable clustering by performing a triad-formation step in addition to growth and 
		preferential attachment (which are the same as in the \emph{BA model}~\cite{barabasi1999emergence}).
        
		Thus, we consider two levels of mean clustering for scale-free networks (\emph{low} and 
		\emph{high}) by tuning a simple parameter~\cite{holme2002growing}.
		The small-world property is not a binary one, and as such, there exists a degree of what we would
		call \emph{small-world-ness}.
		The process of tuning the mean clustering coefficient in our scale-free networks has an 
		immediate effect in the degree of small-world-ness of such structures.
		 
		\subsection{Broadcasting hubs and absorbing hubs} 
            In directed networks, the in-degree and the out-degree distribution are generally not the same.
			Some directed networks possess a power-law distribution in both their in-degree and out-degree distributions, 
			but with different exponents (e.g.~the WWW~\cite{newman2003structure}), whereas others have a power law only in
		 	one of the two directions (e.g.~citation networks~\cite{newman2003structure}).
		 	In either case, the presence of a long-tail in the \emph{out-degree} distribution of a 
		 	network implies the existence of \emph{broadcasting hubs}, that is, nodes that have massive 
		 	outgoing connections compared with other nodes in the system. 
		 	On the contrary, the presence of a long-tail in the \emph{in-degree} distribution
		 	implies the existence of \emph{absorbing hubs}, that is nodes that have a big amount of 
		 	incoming connections.
		 	Here, we are interested in analysing how collective dynamics develop for the case of networks
		 	with broadcasting hubs and for networks with absorbing hubs.

		\subsection{Spike-timing dependent plasticity (STDP)}
			\label{model:STDP}
			As mentioned in Sect.~\ref{intro}, STDP is a temporally asymmetric form of Hebbian learning 
			induced by temporal correlations between pre- and postsynaptic neurons. Synaptic
			weight between pre- and postsynaptic nodes is potentiated (increased), 
			if the postsynaptic neuron fires shortly after the presynaptic neuron. 
			It is depressed 
			(decreased) if the opposite happens, namely the post-synaptic 
			neuron fires shortly \emph{before} the pre-synaptic neuron.
            We implemented STDP mechanisms in our model through the following set of equations:
			\begin{equation}
			    \label{eq:STDPupdate}
				w_{ij}(t+1) = w_{ij}(t) + \Delta w_{ij}(\Delta t) 
			\end{equation}
			\noindent where $\Delta t = t_{post} - t_{pre}$ denotes the difference between spikes of pre-
			and post-synaptic neurons, and:
			\[ \Delta w_{ij}(\Delta t) = \left\{
			\begin{array}{l l}
				a_{p} \exp\{\frac{-\Delta t}{T_{p}}\} & \quad \text{if $\Delta t\geq 0$}\\
				-a_{d} \exp\{\frac{\Delta t}{T_{d}}\} & \quad \text{if $\Delta t< 0$}
			\end{array} \right.\]		
		
			\noindent where parameters $a_{p}$ and $T_{p}$ set the amount and duration of LTP, whereas 
			$a_{d}$ and $T_{d}$ set the amount and duration of LTD.
			In our experiments we set $a_{p} = a_{d} = 0.1$.
			Observations of STDP in brain tissue suggest that the time window for potentiation is 
			typically shorter than the depression time window~\cite{bi1998synaptic}, 
			for that reason we let $T_{p} = 10$ time-steps and $T_{d} = 20$ time-steps.
			However, it also has been observed that time-windows and amount of potentiation/depression 
			vary across species and brain structures~\cite{bi2001synaptic}.
		
			We impose hard bounds on synaptic weights, that is,
		    $0<w_{min}<w_{ij}<w_{max}$ $\forall i, j$, which prevents unbounded weight growth, gives
			rise to strong competition between inputs to a neuron and results in a bimodal distribution 
			of the synaptic weights at the end of simulation time~\cite{billings2009memory}.
			In order to allow for activity-dependent pruning of synapses, we set 
			$w_{ij} = 0$, if $w_{ij}\leq w_{min}$ following application of Eq. \ref{eq:STDPupdate}. If the 
			connection is to be terminally deleted, we set also $A_{ij}=0$ once $w_{ij} = 0$.

		\subsection{Estimation of small-world-ness.}
		    \label{model:small-world}
			Following the ideas in Ref.~\cite{humphries2008network}, we estimate the degree of 
			small-world-ness in a network through the following process.
			Let $G$ be a network consisting of $n$ nodes and $e$ edges. To test whether $G$ exhibits the 
			small-world property we construct a random network $R$ with same number of nodes and same 
			number of edges.
			Then, we estimate the mean clustering coefficient of both networks $CC_{G}$ and $CC_{R}$ 
			along with their mean path length $L_{G}$ and $L_{R}$.
			Finally, we compute $S$ as the ratio of those values:
			\begin{equation}
				S = \frac{CC_{G}/CC_{R}}{L_{G}/L_{R}}
				\label{S_SW}
			\end{equation}
			If $S > 1$, then $G$ possesses the small-world property~\cite{humphries2008network},
			which implies a more abundant presence of cliques and long-range connections among nodes 
			than expected by chance.

%
		
		\subsection{Numerical implementation}
		    \label{model:implementation}
			When starting simulations, all membrane potentials are initialised at random, whereas states 
			are set to inactive. The neural threshold $\theta$ will always be set to unity.
			We let the system reach the critical state, which is identified by a power-law approximation 
			of the distribution of avalanches (see below), which implies that large events coexist with 
			small events during running time. 
            Afterwards we apply STDP mechanisms to update the synaptic strength among nodes based on 
            their activity. At the end of simulations, we analyse the resulting network and the dynamics 
            of the system.
		
		    Both the relaxation time towards the critical state as well as the sampling time needed to 
		    assess criticality depend on the system size. We consider network sizes $N = 128$, $256$ and $512$ and
		    choose the initial `settling' time of the networks to, resp., $10^6$, $2\times 10^6$, $3 \times 10^6$ 
		    before introducing STDP mechanisms in the system.
			This selection of times is appropriate for large events to take place during simulation time.
		
			Results are fitted by power-law distributions and the quality of the fit is evaluated by 
			the mean-squared deviation $\Delta\gamma$. This is obtained based on the best-matching power-law 
			exponent $\gamma$ from linear regression in log-log scales.
			Our choice of using this method is due to its simplicity and justified by the asymptotic 
			unbiasedness of the estimation.
			When this error function is at its minimum, that is, when the data is best approximated by a 
			power-law distribution with exponent $\gamma$, is then when the system is at criticality.

		    Moreover, following Ref.~\cite{larremore2011predicting} we inspect the value of the largest eigenvalue
		    of the matrices $W$ associated to each network and whose entries $w_{ij}$ denote the synaptic weight
		    between node $i$ and $j$.
		    The authors in Ref.~\cite{larremore2011predicting} observe that the largest eigenvalue of the weight matrix
		    governs the dynamics of the system.
		    Through an analytical derivation, it is reported that when the largest eigenvalue equals unity the system is at the
		    critical state.

	In Table~\ref{tabLargstEig} we report the value of the largest eigenvalue $\Lambda$ of the weight matrices associated to our networks.
	In our experiments the critical state is not only identified by the power-law distribution of avalanche sizes (see Fig.~\ref{fig3}) before the onset of STDP mechanisms,
	but also by the value close to unity of $\Lambda$.

			After the system has reached such a state, we let STDP mechanisms to set in at the synaptic 
			level, that is, at the level of individual coupling strengths.
			For all system sizes and topologies we allow four million time steps of STDP regime, after 
			which the simulation is over and we proceed to analyse the structural
			changes of the system.
			Unlike the model in Ref.~\cite{basalyga2011emergence} we do not distinguish between STDP-neurons 
			and non-STDP-neurons, rather in our model every synapse in every unit is susceptible of 
			plasticity.
		
			For our experiments we consider $50$ different networks per class (\emph{b} to \emph{d} in Sect.~\ref{Sect:model}) and system 
			size for the sake of statistical robustness. 
			In the case of fully-connected networks, as there exists only one fully-connected network of 
			size $N$, randomness is introduced in our code for each realisation of the experiment rather 
			than in the structure as for the other network classes considered.
			After each realisation of the experiment we generate $100$ random networks to compute the 
			metric $S$ (see Sect.~\ref{model:small-world}) for each of the 
			networks that result
			from the simulation.
			Experiments were carried out in the EDDIE computer cluster of the University of 
			Edinburgh.

	\section{Results}
		\label{results}
		
		\subsection{Small-world structure emerges in fully-connected networks}
\begin{figure}[t]
				\centering
				\subfloat[Fully-connected]{%
					\includegraphics[scale=0.28]{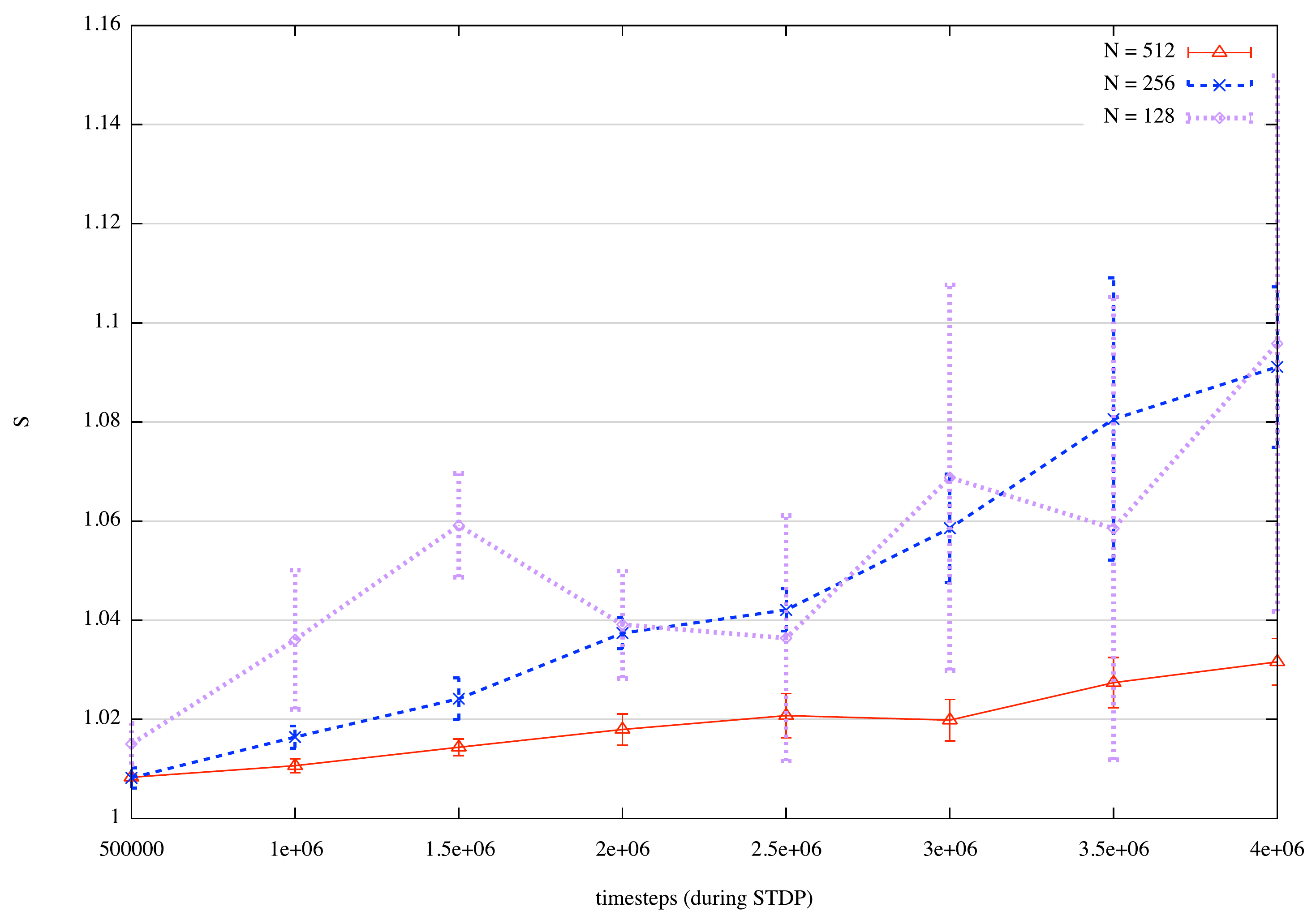}
					\label{fig1:a}
				}\\
				\subfloat[Random]{%
					\includegraphics[scale=0.28]{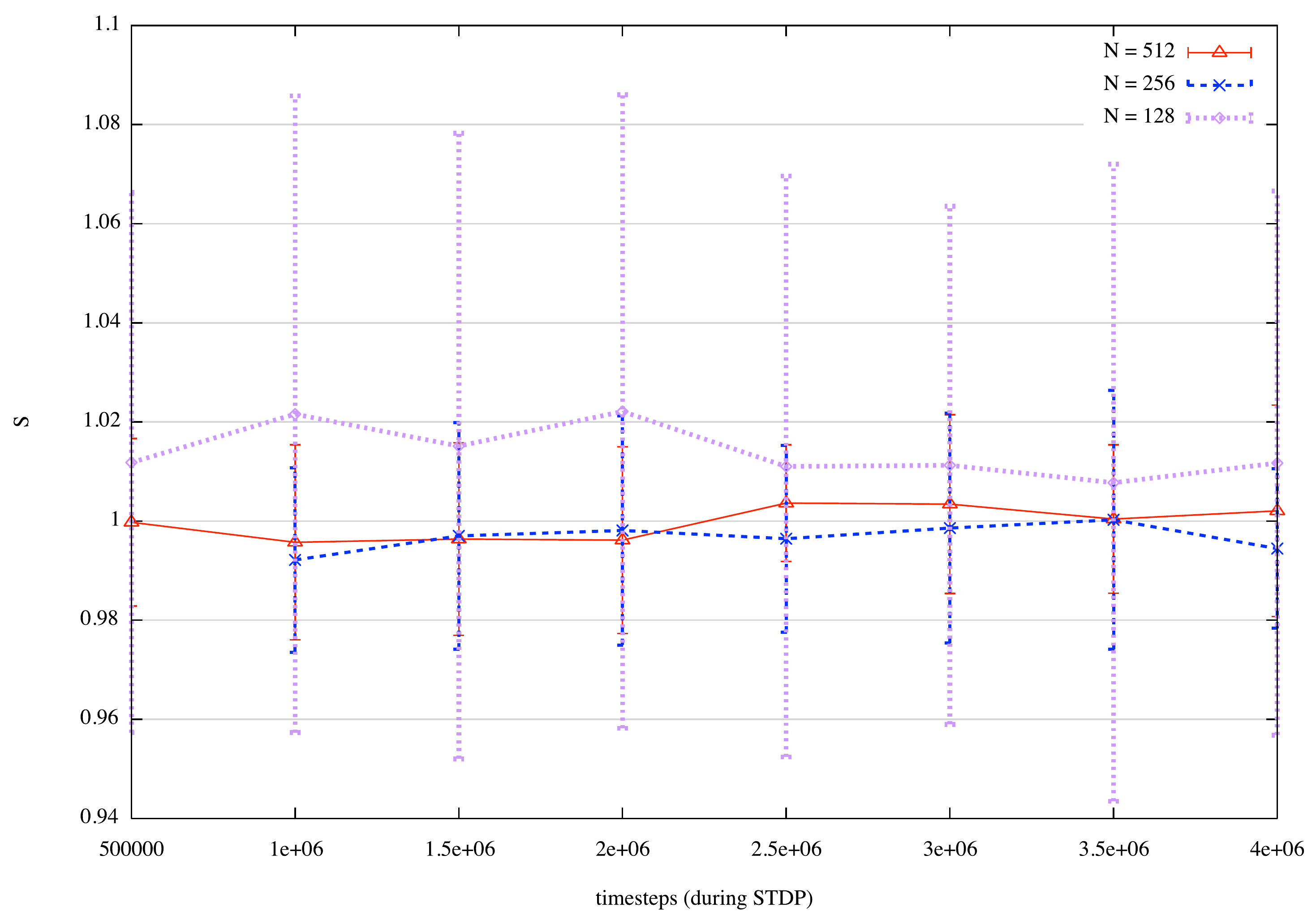}
					\label{fig1:b}
				}
				\caption{
				    Evolution of small-world-ness measured by Eq.~\ref{S_SW}
				    for fully-connected (a) and random (b) networks.
				    Small-world-ness increases due to STDP
				    for fully-connected networks, which
				    is not observed for the random networks.
				    This more evident for smaller system sizes.
				    We show mean values with error bars representing standard deviations.
				}
				\label{fig1}
			\end{figure}

			We observe that after simulation time a small-world structure emerges from fully-connected 
			networks as measured by the metric $S$.
			In Fig.~\ref{fig1} we show the evolution of $S$ per time step for the fully-connected and 
			random topologies considered.
			(We present mean values and standard deviations estimated from the different realisations of 
			our experiment.)
			
			Unlike the fully-connected case, the random structure does not show any particular trend 
			regarding the evolution of $S$.
			Therefore, STDP does not imply any improvement in the structure of random networks.
			Moreover, as we can see in Fig.~\ref{fig1}, STDP acts faster in smaller systems than in 
			larger ones, thus its effect is clearer in the networks comprising
			$128$ nodes than in those of $512$ nodes. In any case, we observe a positive trend in the 
			evolution of $S$.

			What is the effect of STDP in the scale-free networks considered?
			As mentioned previously, these structures already possess the small-world topology, that is 
			$S > 1$ for all of them, however the degree of \emph{small-world-ness}
			in them vary. We considered \emph{high} and \emph{low} values of the mean clustering 
			coefficient for this type of networks, this results respectively in high 
			and low degrees of small-world-ness in the networks considered. The reason we considered such
			classes is because we are interested in assessing how STDP 
			mechanisms affect scale-free structures with varying clustering coefficients.
			Additionally, as stated previously, we consider scale-free networks with 
			\emph{broadcasting hubs} and networks with \emph{absorbing hubs}.
		
\begin{figure}[t]
				\centering
				\subfloat[Low CC]{%
					\includegraphics[scale=0.28]{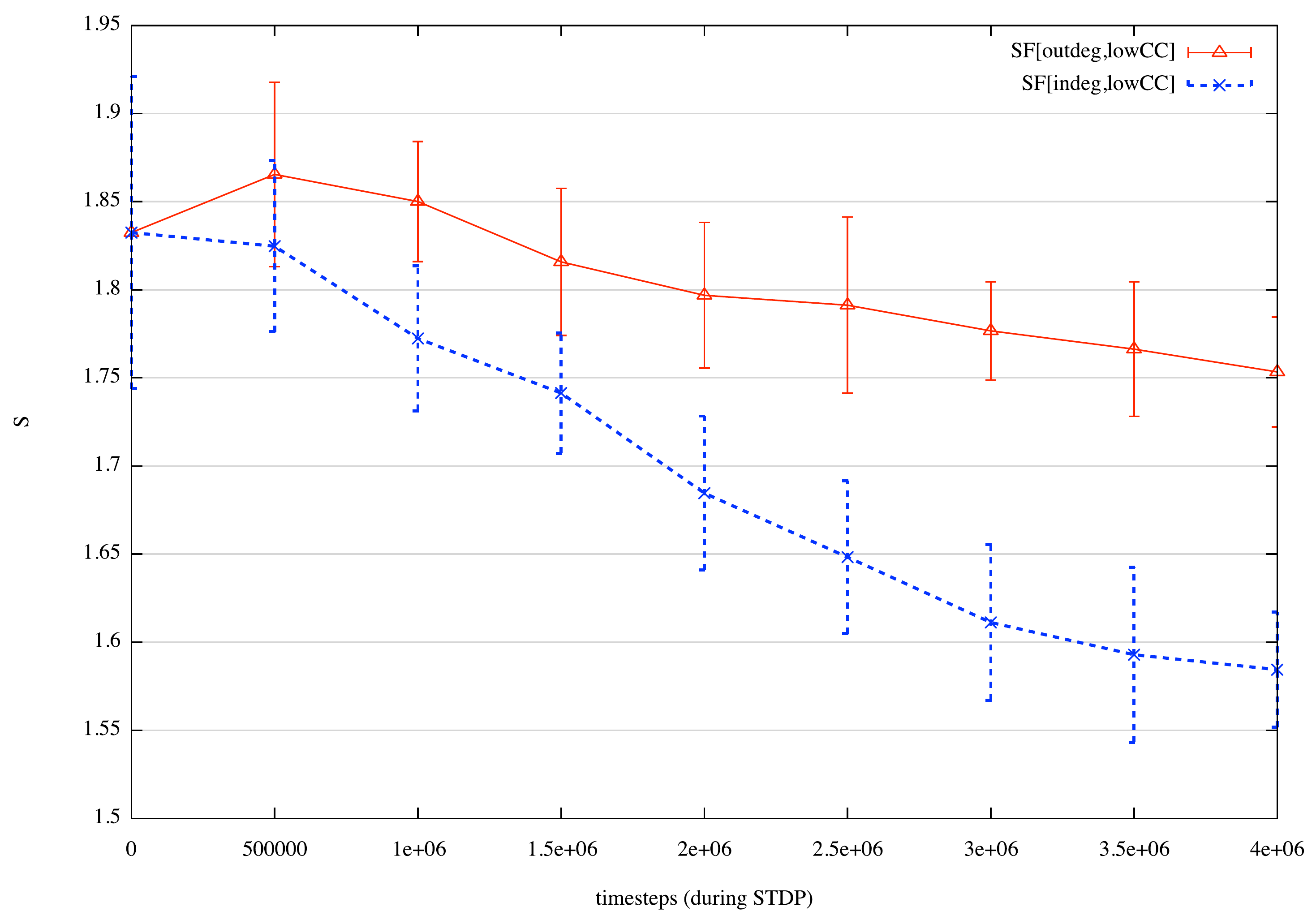}
					\label{fig2:a}
				}\\
				\subfloat[High CC]{%
					\includegraphics[scale=0.28]{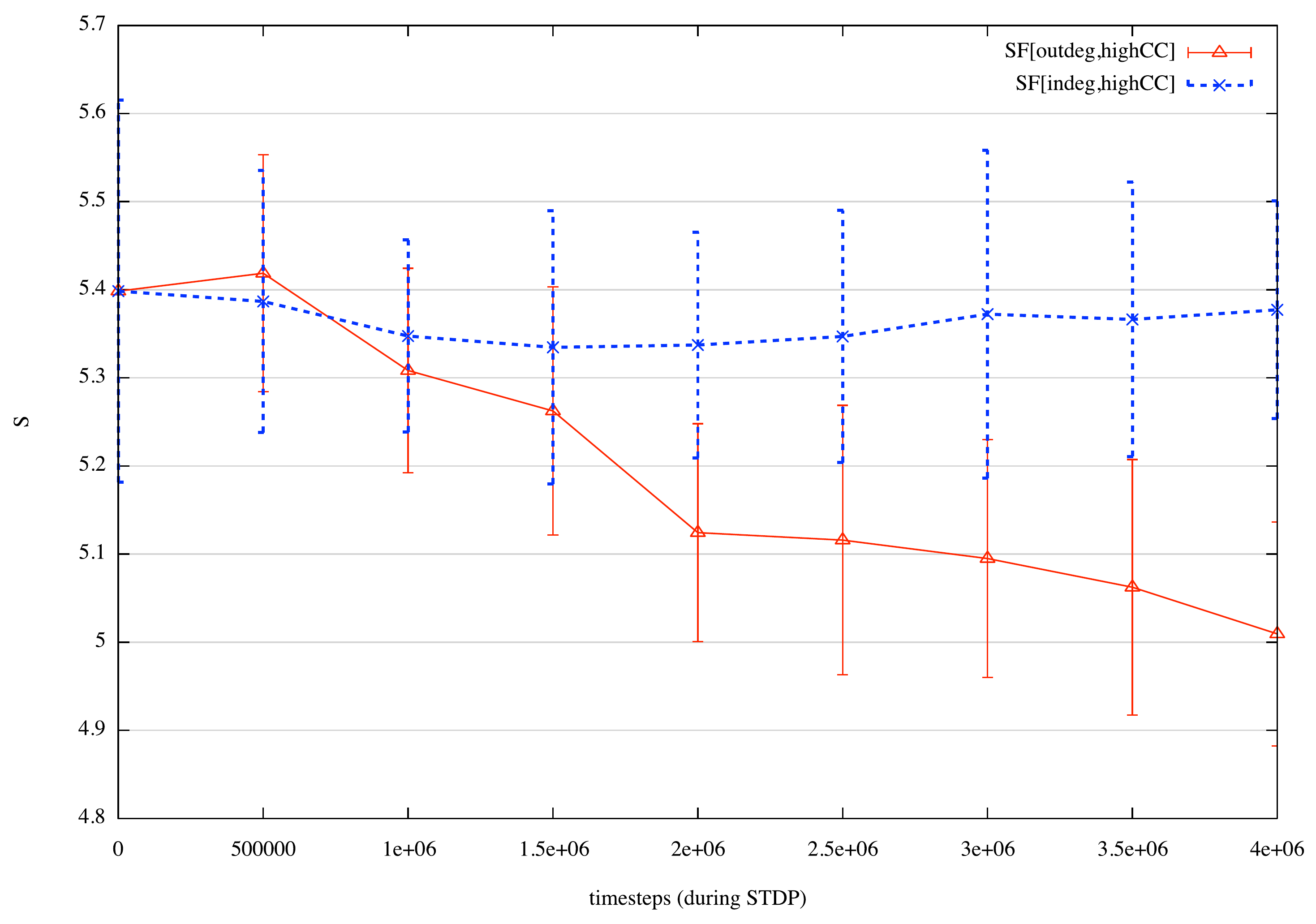}
					\label{fig2:b}
				}
				\caption{
				    Effects of STDP on scale-free networks of size $N = 512$.
				    STDP reduces the degree of small-world-ness in scale-free
				    networks. This is observed across all system sizes and when
				    the network has either low (a) or high (b) mean clustering coefficient.
				    We compare the cases when the power law distribution is either in the
				    out-degree (continuous line) or in the in-degree (dashed line) distribution.
				}
				\label{fig2}
			\end{figure}

\begin{figure}[t]
				\centering
				\subfloat[Distribution of avalanche sizes $P(S)$ 
				    in log-log scales before and after STDP]{%
					\includegraphics[scale=0.28]{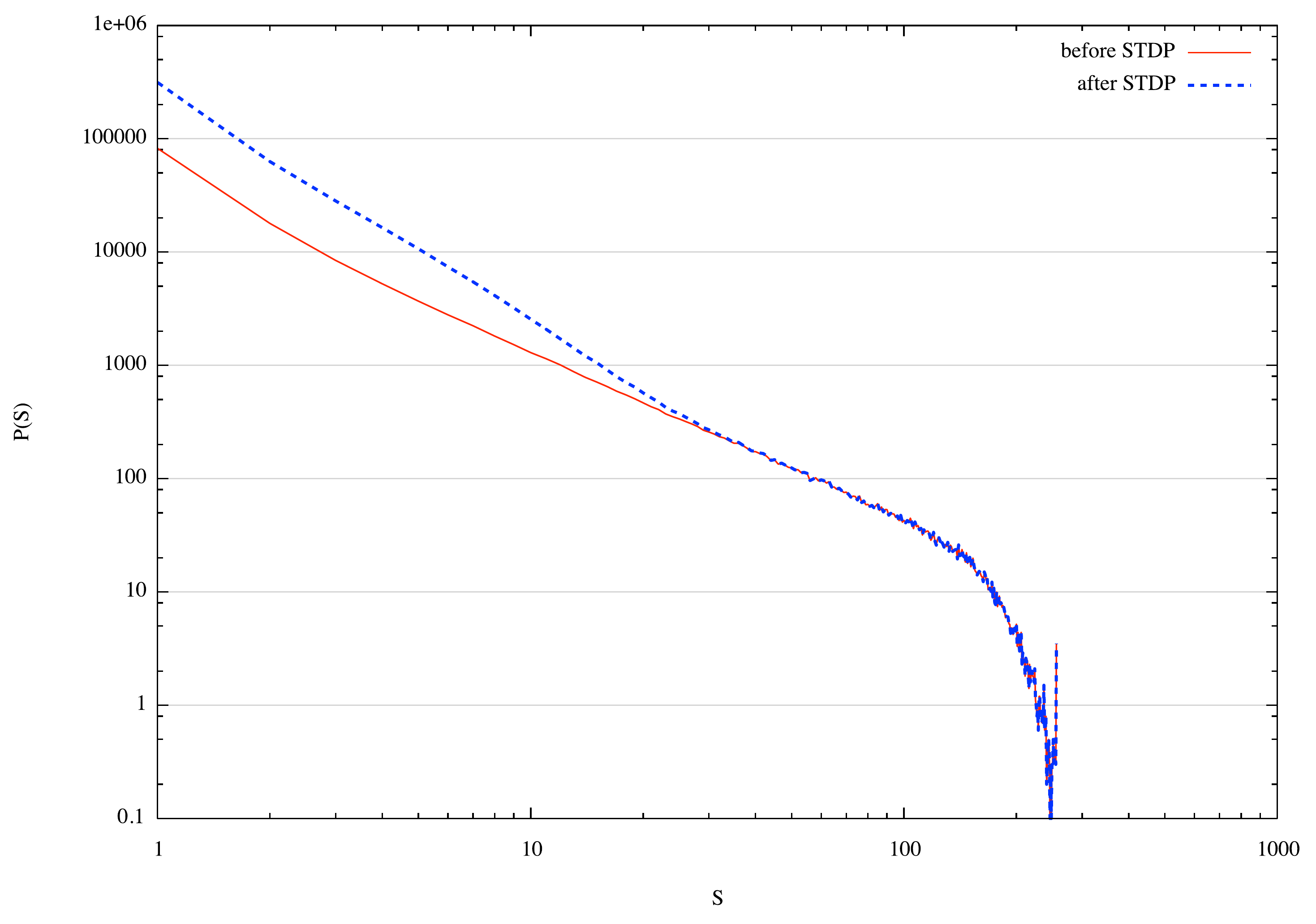}
				}\\
				\subfloat[Deviation from best-fit power-law per time-step]{%
					\includegraphics[scale=0.28]{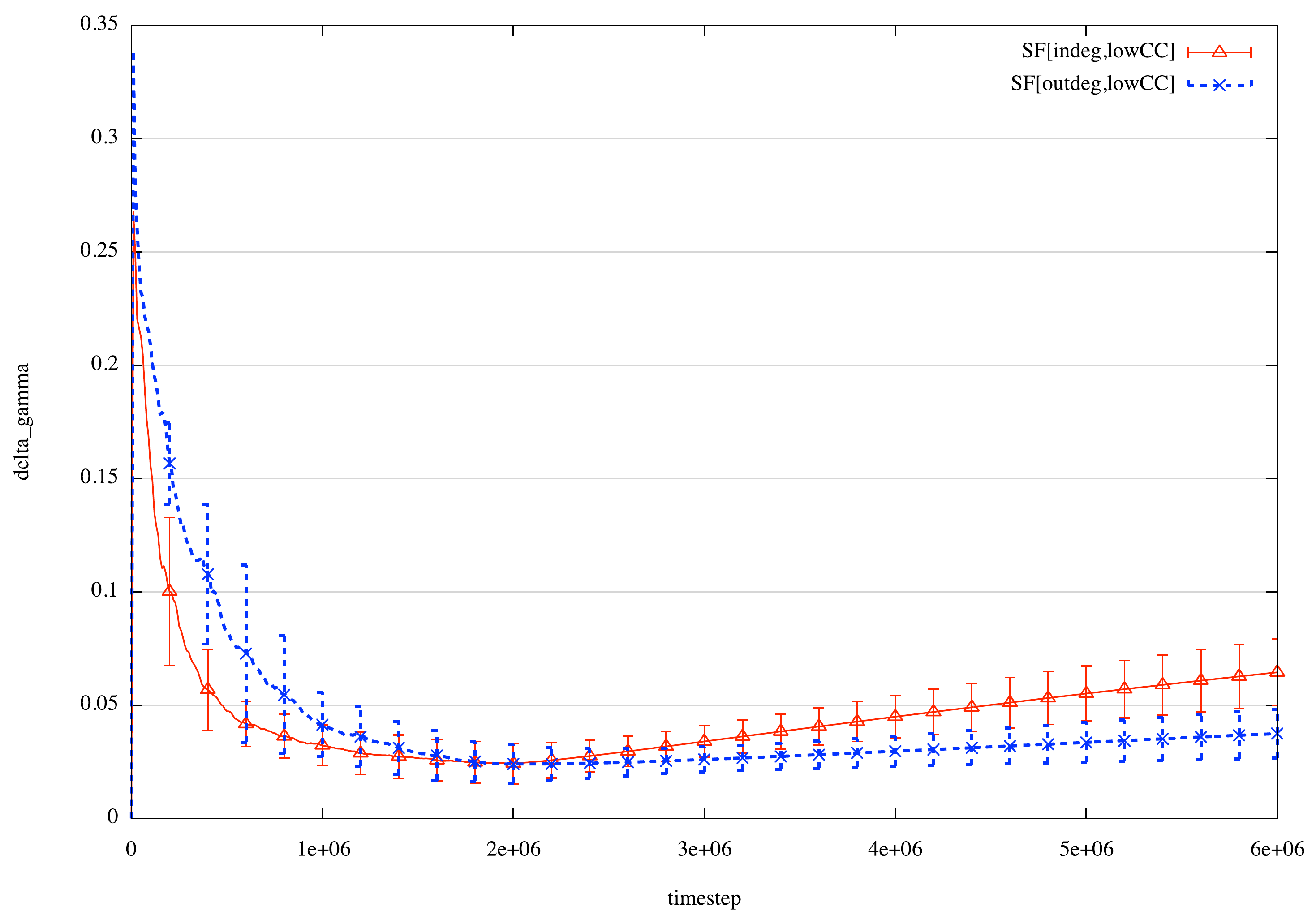}
				}
				\caption{
				    Effects of STDP on collective dynamics for a scale-free net with $N = 256$.
				    Before STDP, small avalanches coexist with large ones, and their distribution
				    can be approximated accurately by a power-law (a, continuous line), which
				    is also identified by a small error around time-step $2\times 10^{6}$
				    (b). 
				    However, after STDP, large avalanches cease to occur (a, dashed line) and
				    the error exhibits an incremental trend for the rest of the simulation
				    (b).
				}
				\label{fig3}
			\end{figure}

			For these topologies, STDP has a negative effect regarding their small-world-ness, identified
			by a decrease in the value of $S$ once STDP sets in.
			For example, in Fig.~\ref{fig2:a} we show a comparison of the evolution of $S$ for two 
			scale-free networks with $512$ nodes and low mean clustering
			coefficient (which implies low degree of small-world-ness). The network identified by the 
			continuous line has a power-law out-degree distribution (which
			yields broadcasting hubs), whereas the dashed line represents its transpose, that is, a 
			network with a power-law in-degree distribution (which yields
			absorbing hubs). As it can be seen, STDP affects negatively the small-world property in these
			networks.
		
			Similarly, Fig.~\ref{fig2:b} traces the changes of $S$ for two scale-free networks of same 
			size as above but with high degree of small-world-ness.
			The continuous line represents a scale-free network with power-law out-degree distribution, 
			whereas the dashed line represents its transpose, a network with 
			a power-law in-degree distribution. In the best of cases, $S$ does not exhibit an noticeable
			decrease.
			These situations are verified in all other systems sizes considered for scale-free networks.
		
			It is clear that STDP does not have the same effects when the power-law of the degree 
			distribution is present in the out-degree distribution 
			or in the in-degree distribution; nor when the clustering coefficient of network is low or high. 
			The trend, however, is not unambiguous. Nevertheless, we find 
			that STDP affects negatively the degree of small-world-ness 
			in scale-free networks.
		
		\subsection{STDP impairs criticality}
			As mentioned in Sect.~\ref{model:implementation}, 
			we assess the quality of the power-law approximation to the
			distribution of avalanches by estimating the deviation from the best power-law fit.
			When such an error function reaches a minimum value of less than 0.05, we consider the 
			event-size distribution as well approximated by a power-law and conclude that the system 
			is in a critical state. After the system has approached a power-law behaviour, we allow 
			STDP mechanisms to set in.
			We observe that criticality is generally lessened while STDP is modifying the synaptic 
			strengths. 
			With STDP the difference between the distribution of avalanche sizes and the 
			best-fit power-law increases again, such that eventually a power-law ceases to appropriately
			model the system, see Fig.~\ref{fig2:b}. 
			A similar behaviour as shown there 
		    is observed in all the networks that we considered.
			
			The deterioration of the power-law approximation is explained by the fact that once STDP sets in
			large avalanches cease to occur and only small avalanches take place.
			In this sense, it can be said that STDP favours local, clustered events.
			Fig.~\ref{fig2:a} shows an example of this behaviour. 
			Here, the continuous line shows the system during criticality before STDP mechanisms take 
			place, which can also be identified in the minimum value  shown in Fig.~\ref{fig2:b} around 
			the time step $2\times 10^{6}$. 
			However, after STDP there are no more large avalanches to be added up to the distribution, 
			only small ones, which results in
			the particular shape of the dashed line in Fig.~\ref{fig2:a}.
			As well, in Fig.~\ref{fig2:b} after STDP has set in right after the $2\times 10^{6}$ time 
			step the deviation from a power-law distribution increases and
			we observe a larger error as time passes.
			
			As mentioned before, this same phenomenon is observed across all networks considered in our 
			model.
			This behaviour is captured by the largest eigenvalue of the weight matrix $W$ denoted by $\Lambda$.
			Before STDP the value of $\Lambda$ is close to unity. However, once STDP mechanisms set in, the 
			synaptic modulation results in a weight matrix whose largest eigenvalue is less than unity, thus reflecting
			the deviation from criticality (see Table~\ref{tabLargstEig}).
			From this we might conclude that in our model the critical state vanishes as STDP sets in.
			
			Does this imply that self-organised criticality and spike-timing-dependent plasticity are two biological phenomena
			that cannot coexist?
			The model proposed by Levina et al.~\cite{levina2007dynamical}, based on dynamical synapses, exhibits critical behaviour
			that coexist with STDP.
			The modulation induced by such dynamical synapses results in a compensatory mechanism that recovers the critical regime.
					
		\subsection{STDP prunes direct-feedback connections} 
\begin{figure}[t]
				\centering
				\includegraphics[scale=0.16]{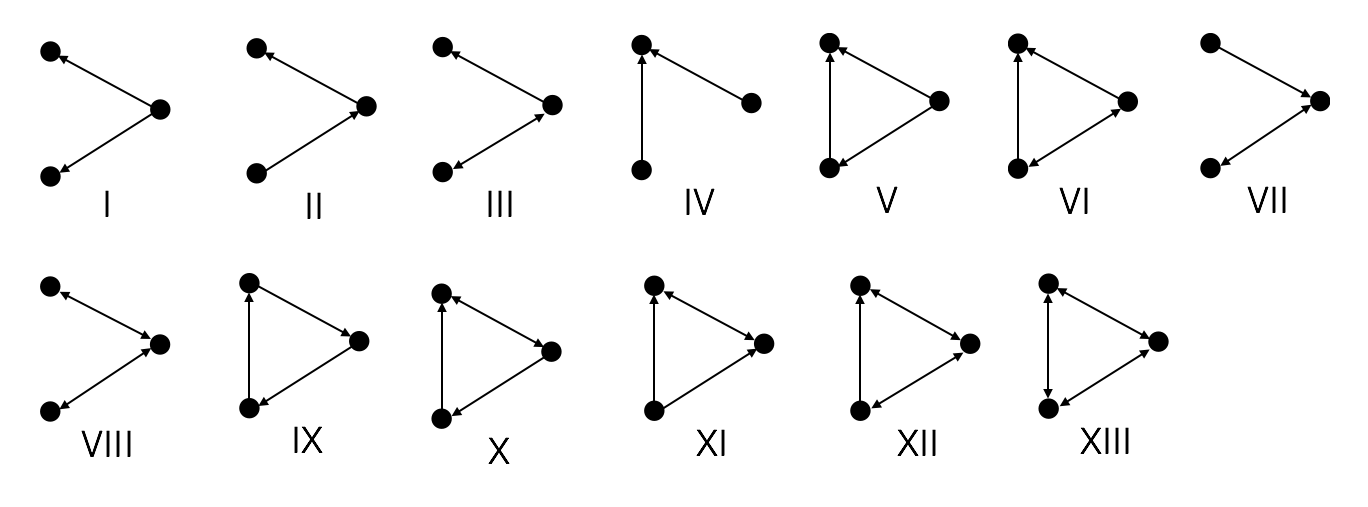}
				\caption{All configurations for $3$-node connected subgraphs.}
				\label{fig4}
			\end{figure}

			We analyse the motif profile of the networks after the STDP regime.
			There are $13$ different motifs representing the possible 
			relations among nodes taken in threes from a 
			directed network. The different configurations are shown in Fig.~\ref{fig4}.

\begin{figure}[t]
				\centering
				\includegraphics[scale=0.28]{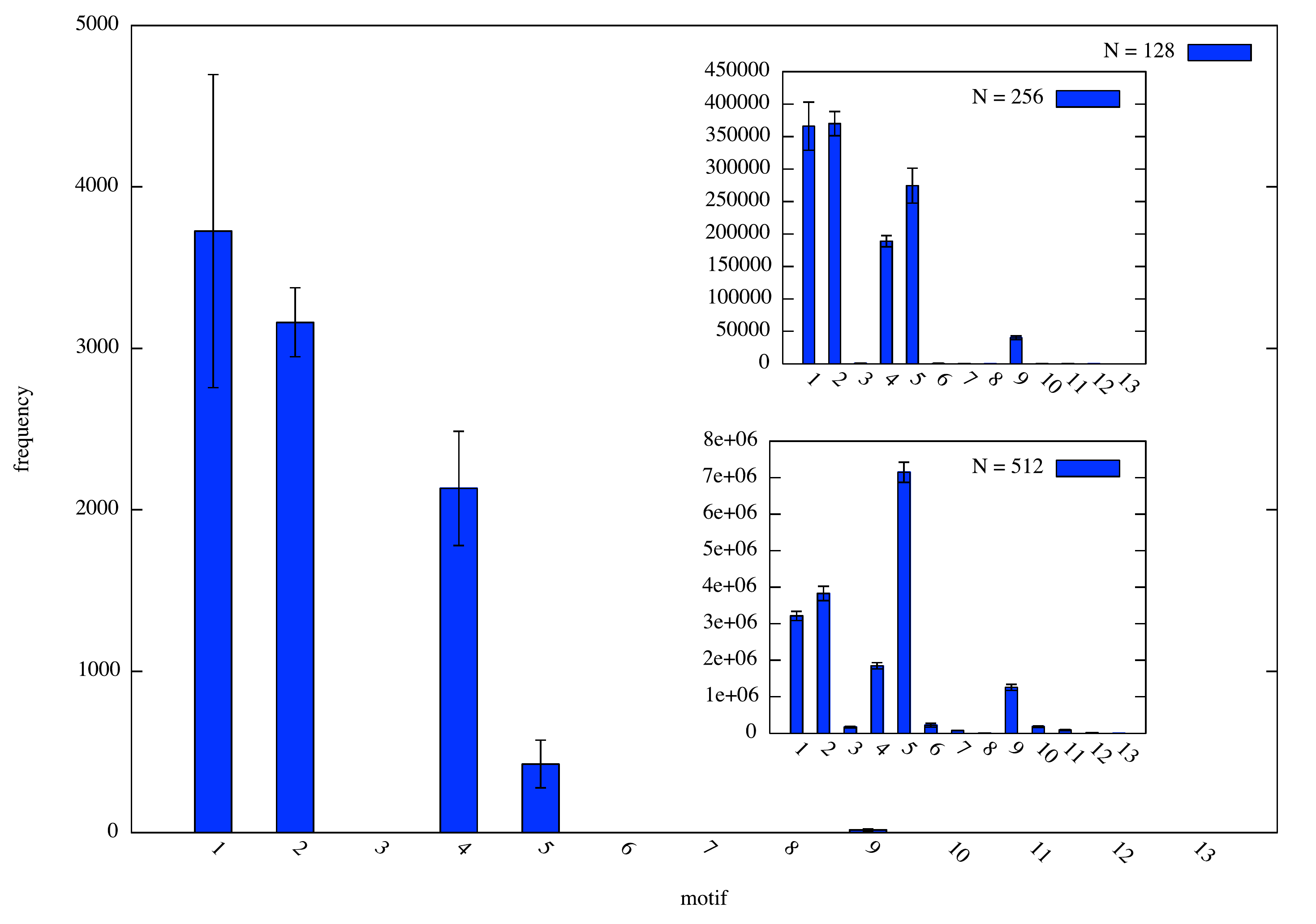}
				\caption{
				    Post-STDP motif profile for fully connected networks of size $N =  128$. 
				    (Inset: sizes $N = 256$ and $N = 512$.)
				    After STDP, motif \emph{XIII} breaks apart and motifs
			        \emph{I}, \emph{II}, \emph{IV}, \emph{V} and \emph{IX} emerge.
			        This is more evident for smaller system sizes, where STDP requires less
			        time for pruning.
				}
				\label{fig5}
			\end{figure}

			A motif profile shows the distribution of these $13$ different $3$-node configurations for a 
			single network.
			For example, before STDP, a fully-connected network possess a single motif distribution: all 
			$3$-node relationships are of the \emph{XIII} type
			(see Fig.~\ref{fig4}).
			However, after STDP the motif \emph{XIII} breaks apart and instead motifs of the types
			\emph{I}, \emph{II}, \emph{IV}, \emph{V} and \emph{IX} abound.
			Interestingly, none of these contain direct feedback connections.
			Fig.~\ref{fig5} shows the motif profiles of the fully-connected networks considered (sizes 
			$256$ and $512$ inside the inset) after STDP. 
			For the smallest system size it is more evident how motifs \emph{I}, \emph{II}, \emph{IV}, 
			\emph{V} and \emph{IX} grow where previously there was only motif \emph{XIII}.
			As the system grows, STDP requires more time to prune the network and
			profiles differ from one another.
									
\begin{figure}[t!]
				\centering
				\subfloat[outdegree-SF net with low CC (inset: its transpose).]{%
					\includegraphics[scale=0.28]{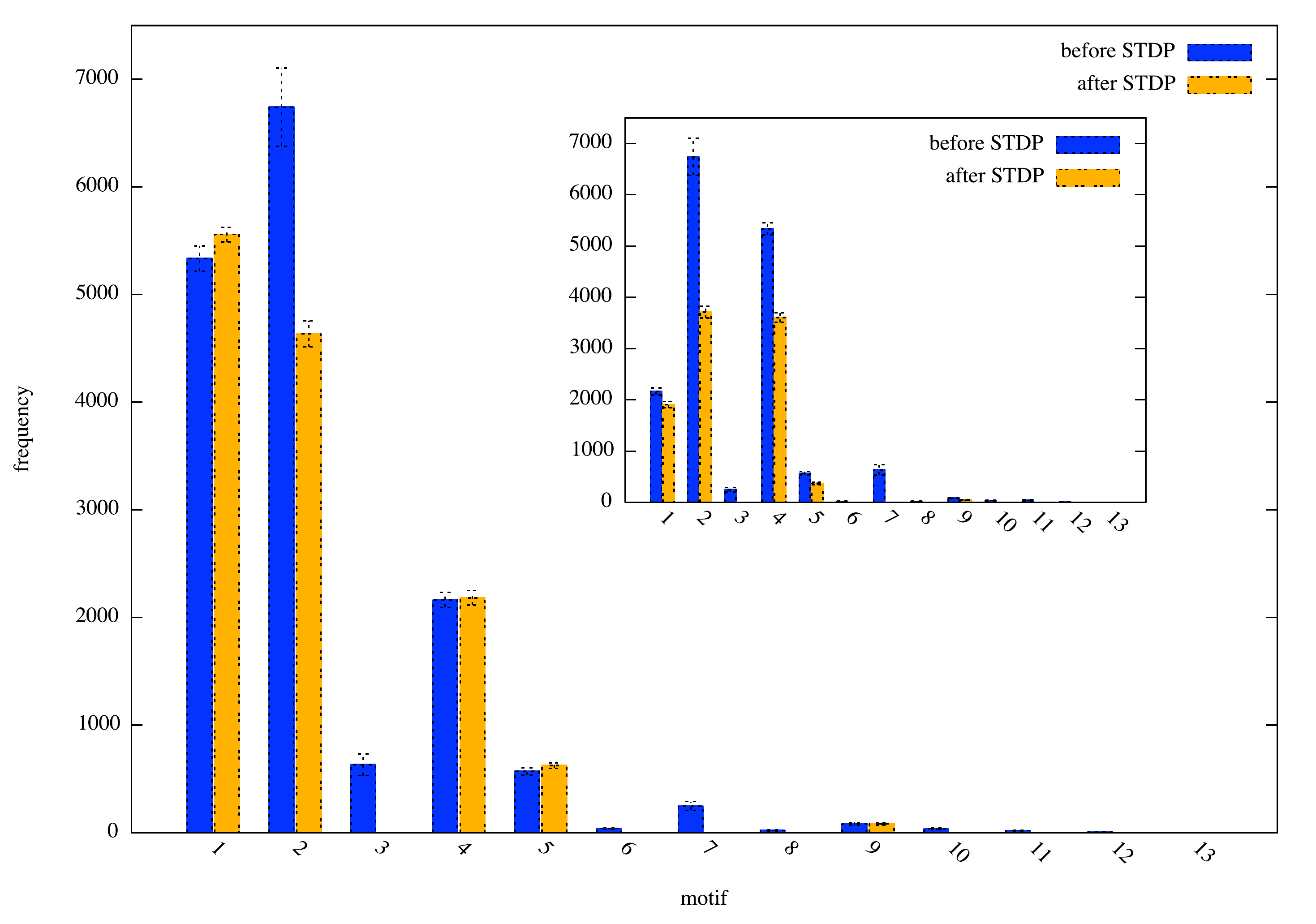}
					\label{fig6:a}
				}\\
				\subfloat[outdegree-SF net with high CC (inset: its transpose).]{%
					\includegraphics[scale=0.28]{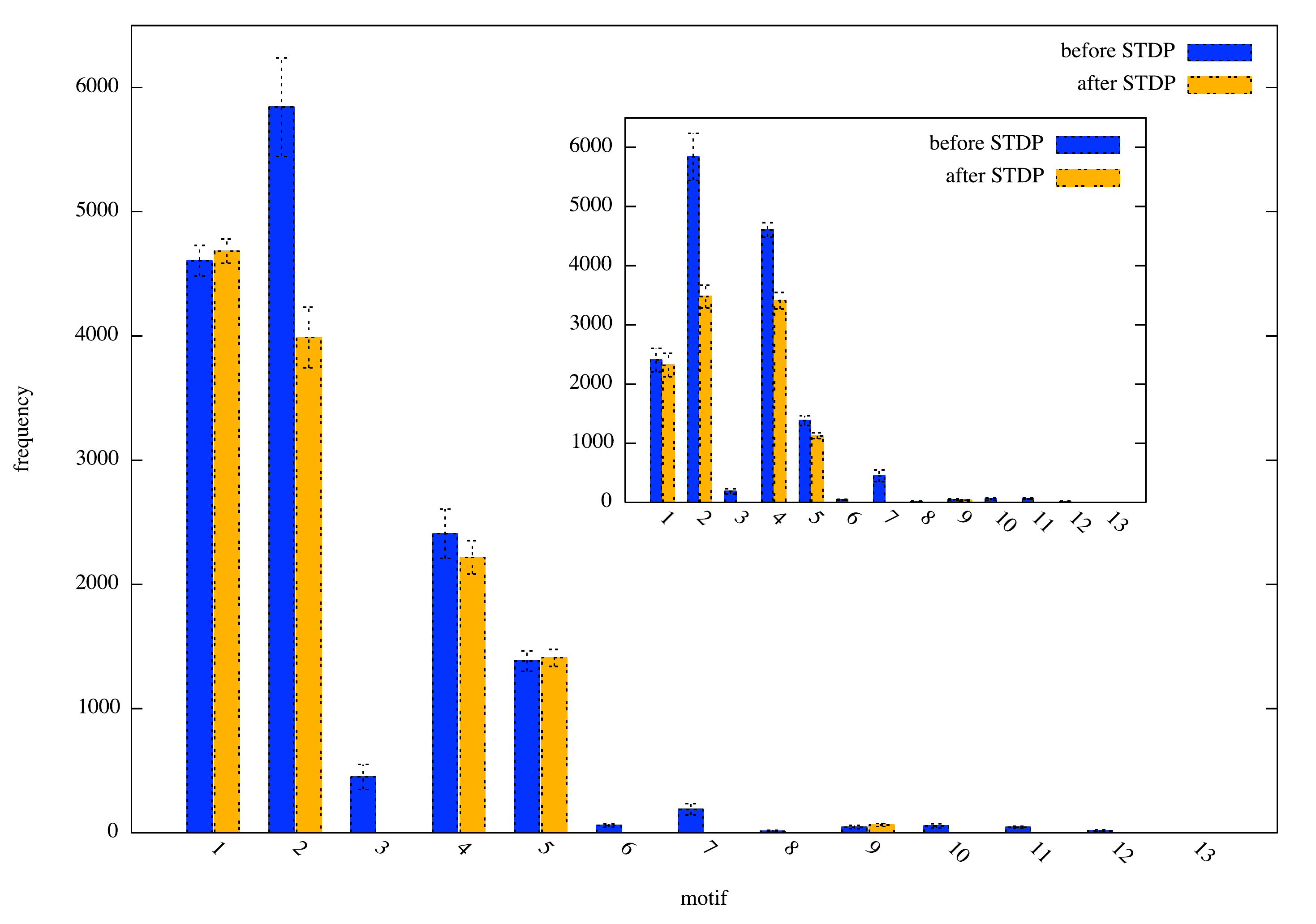}
					\label{fig6:b}
				}\\
				\caption{
				    Motif profiles for networks of size $N = 128$ pre- (blue)
				    and post-STDP (orange) regime.
				    STDP particularly affects motifs that contain direct feed-back connectivity (note that we are not assuming refractoriness in the neurons, otherwise higher-order motifs had to be considered).
				    This is observed across all system sizes and when the network possess 
				    high (a) or low (b) mean clustering coefficient (CC), 
				    as well as when
				    the network has an in-degree power-law distribution (insets).
				    A similar behaviour is observed in random networks (not shown).
				}
				\label{fig6}
			\end{figure}

			In the case of heterogeneous topologies, STDP \emph{attacks} motifs with bidirectional 
			connections affecting in this way the local clustering
			of the global structure, which in turn affects the degree of small-world-ness of the network.
			Motifs with bidirectional connections are: \emph{III}, \emph{VI}, \emph{VII}, \emph{VIII}, 
			\emph{X}, \emph{XI}, \emph{XII} and \emph{XIII}; all of them impaired by the action of STDP.
			Fig.~\ref{fig6} shows this behaviour for our networks of $128$ nodes. However, we observe 
			this particular behaviour in all the different system sizes considered.
			
			In particular, motif \emph{II} is severely impaired in all cases considered.
			This motif represents the most basic feed-forward flow of information comprising $3$ nodes, 
			and as such the most elementary pre- and post-synaptic relationship among three nodes, and as
			such it is expected to be a target for STDP.
			Interestingly, some motifs vanish completely from the profile. These are motifs \emph{III}, 
			\emph{VI}, \emph{VII}, \emph{VIII}, \emph{X}, \emph{XI}, \emph{XII} and \emph{XIII}; all of 
			them involving direct feed-back connectivity.
			
			Therefore, in a sense it can be said that STDP prevents direct feedback connectivity, 
			preferring indirect feedback flow in which a third node serves as intermediary.
			For example, motif \emph{X} includes a direct feedback connection between two nodes; however, 
			after STDP this motif might transform into motif
			\emph{IX}, in which the feedback flow is now mediated by a third node.
			
			STDP favours \emph{one-way} connections rather than \emph{two-way} connections, thus 
			establishing a direction for the flow of stimuli given by the current activity in the system.
			Simply put, if node $i$ sends an edge to node $j$, but this latter node fires shortly before 
			the former most of the time, then STDP acts by severing this somewhat \emph{erroneous} 
			connection.
			
		\subsection{Other effects on topology} 
			Previously, we mentioned how STDP affects the degree of small-world-ness in the networks 
			considered.
			The process of activity-dependent pruning has other consequences in topology, namely, the 
			decrease in edge density and eventually, the disconnection of the network and the emergence 
			of different connected components across the system.
			
			The density of a network is the ratio of the number of edges to the number of possible edges.
			As such, a fully-connected network possess all possible connections in the network resulting 
			in the maximum possible value of the density, which is $1$.
			As mentioned before, for any of the system sizes considered, the heterogeneous structures 
			have the same number of edges, which yields the same density for all of them.
			However, how STDP affects this ratio varies depending on the particular type of network 
			considered (fully-connected, random or scale-free).
						
\begin{figure}[t]
				\centering
				\includegraphics[scale=0.28]{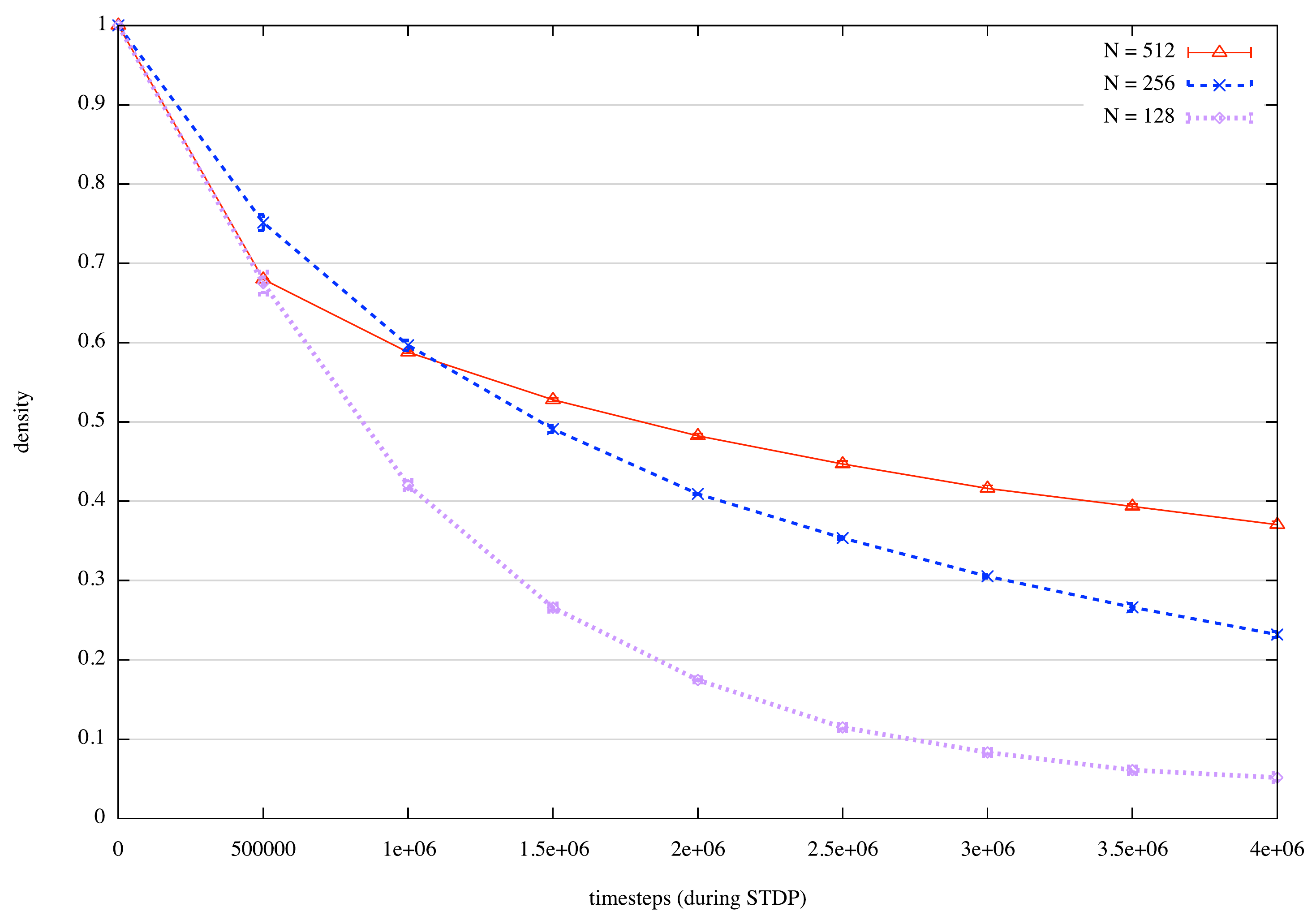}
				\caption{
				    Density of fully-connected nets.
				    STDP prunes edges here faster than in the
				    other structures (see Fig.~\ref{fig7}).
				}
				\label{fig7}
			\end{figure}

			Network density is an indicator of sparseness in a structure. Our heterogeneous networks are 
			sparse, a feature that has an effect on the way STDP acts upon the structure.
			Fig.~\ref{fig7} shows the effects of STDP mechanisms on network density for all the 
			fully-connected cases considered and for the heterogeneous structures of size $128$.
			We observe the same behaviour for the other system sizes considered.
			
			\begin{figure}[t!]
				\centering
				\subfloat[Network density]{%
					\includegraphics[scale=0.28]{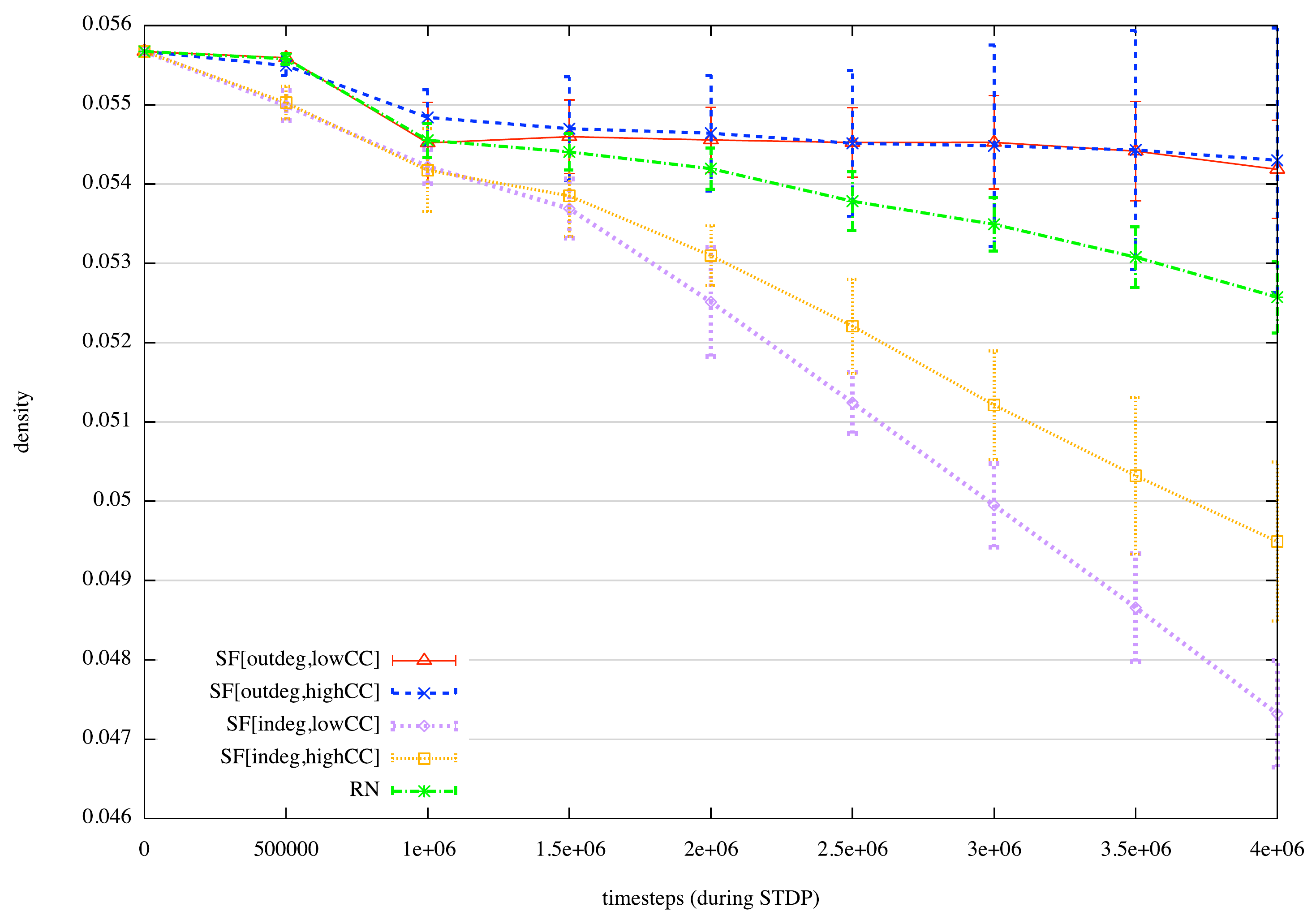}
					\label{fig8:b}
				}\\				
				\subfloat[LCC size]{%
					\includegraphics[scale=0.28]{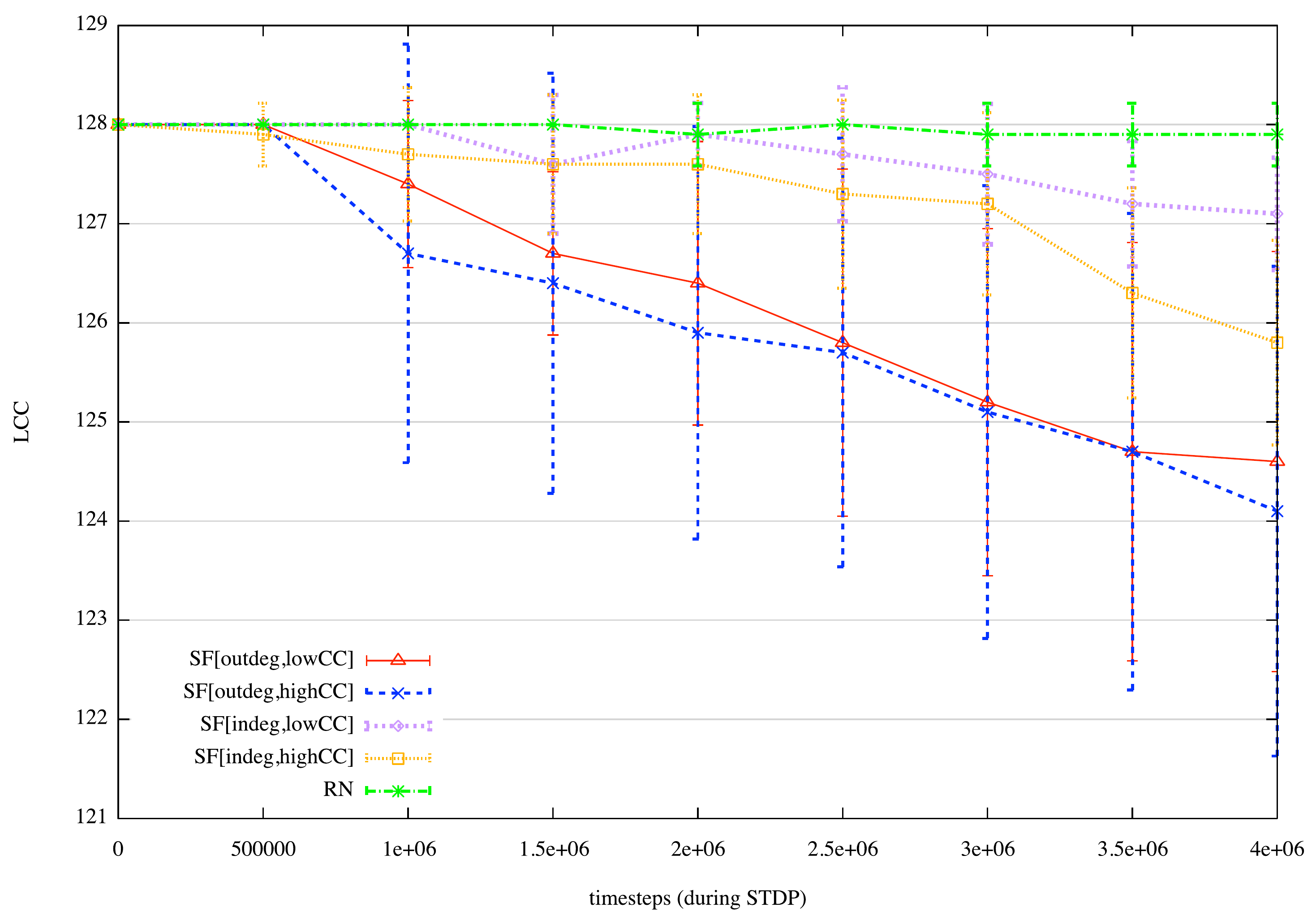}
					\label{fig8:c}
				}				
				\caption{
				    STDP effects on network density and the size of the largest connected 
                    component (LCC) for heterogeneous structures of size $N = 128$.
                    Unlike the fully-connected case, STDP prunes edges slowly in our
                    heterogeneous structures, not even removing $1\%$ of their edges
                    by the end of the experiment.
                    This can be explained by the sparseness of these networks.
                    In-degree scale-free networks with low mean clustering coefficient
                    (CC) lose edges faster than any other network (a), whereas
                    the LCC of out-degree scale-free networks with high CC shrinks faster
                    than any other structure.
                    We observe this behavior across all system sizes.
				}
				\label{fig8}
			\end{figure}	
			
			The effects of STDP are soon visible in a dense network topology. In a fully-connected 
			network,  
			STDP acts fast removing a considerable amount of edges as soon as it sets in.
			In contrast, for sparse-networks, either scale-free or random, STDP acts more slowly not
			even removing $1\%$ of the initial configuration under otherwise similar conditions.
			
			However, STDP does not act the same in every heterogeneous topology. Scale-free networks in 
			which the power-law is present in the in-degree distribution
			(that is, networks with absorbing hubs) lose density faster than any other, specially when 
			the mean clustering coefficient is low.
			This can be explained by the fact that in in-degree scale-free networks absorbing hubs are 
			more susceptible to fire, which in turn has a strong influence on STDP mechanisms. 
			However, clustering seems to serve as a protective mechanism to prevent edge pruning. 
			This can be observed in the fact that our scale-free networks with low mean clustering 
			coefficients lose edges faster than their high clustered counterparts.
			
			Nevertheless, out-degree scale-free networks (that is, networks with broadcasting hubs) 
			decrease the size of their largest connected component faster
			than any other topology, specially when combined with high mean clustering coefficient.
			
			During simulation time, the effects of STDP pruning result on a disconnected network where 
			components of different sizes emerge.
			All our observations regarding topology refer to the largest connected component (LCC) of the
			network.
			Because STDP behaves differently across the different topologies considered, we expect the 
			LCC to be different as well.
			As we have just mentioned, the LCC of scale-free networks with broadcasting hubs suffer a 
			shrinkage of size faster than random networks and
			scale-free networks with absorbing hubs, even if it is this latter type of networks the one 
			that loses density faster than any other topology.
			This behaviour is shown in Fig.~\ref{fig8:c} and it is observed across all system sizes 
			considered.
			
			In summary, what is happening is that scale-free networks with absorbing hubs lose edges 
			faster than any other structure. However, these networks
			show more resilience to become disconnected (Fig.~\ref{fig8:b}). 
			On the contrary, scale-free networks with broadcasting hubs show more resilience to lose 
			edges, but they are more prone to become disconnected when losing edges (Fig.~\ref{fig8:c}).
			
		\subsection{Effects of spike triplets} 
		
			In this section we briefly discuss the effects of extending the implementation of STDP mechanisms
			from pairs of spikes to triplets of spikes.
			The computational implementation of STDP described in Sect.~\ref{model:STDP} takes into account
			pairs of spikes to estimate the synaptic modulation at each connection, namely, one from the
			pre-synaptic node and another from the post-synaptic node, whose order determine the nature of the
			modulation. In the following, we refer to paired-wise STDP as pSTDP.
			
		    It has been reported that pSTDP fails to replicate observations in experimental data.
		    Intuitively, a pre-post pairing followed by a post-pre pairing of the same magnitude would cancel out any
		    synaptic modulation triggered, however this is not what it has been observed in biological experiments.
		    In experiments, models of pSTDP are not able to explain the synaptic modulation triggered by higher-order plasticity rules
		    such as triplets and quadruplets of spikes~\citep{pfister2006triplets,clopath2010voltage}.
		    For this reason, pSTDP has been extended in order to consider \emph{triplets} of spikes rather
		    than just pairs of them.
		    In the following, we refer to triplet-wise STDP as tSTDP to differentiate it from pSTDP.

            A triplet rule for tSTDP involves sets of three spikes: two pre- and one postsynaptic, or one pre- and
            two postsynaptic spikes. With a triplet rule of this form it is possible to fit experimental data from visual
            cortical slices as well as from hippocampal cultures.
            Interestingly, when this rule is based on Poisson spike trains, the learning rule can be mapped to a 
            Bienestock-Cooper-Munro (BCM) learning rule~\citep{pfister2006triplets, gjorgjieva2011triplet}. 
            Moreover, it has been proposed that the triplet rule for STDP is a mechanism used by neuronal networks to perform
            computations that resemble those of independent component analysis (ICA)~\citep{gjorgjieva2011triplet}.

            When we consider triplets of spikes for STDP, we might feel inclined to extend the model further in order to consider
            higher-order STDP rules.
            However, experiments show that spike triplets are able to reproduce data generated by higher-order 
            terms~\citep{pfister2006triplets}.
		    
\begin{figure}[t]
				\centering
				\includegraphics[scale=0.3]{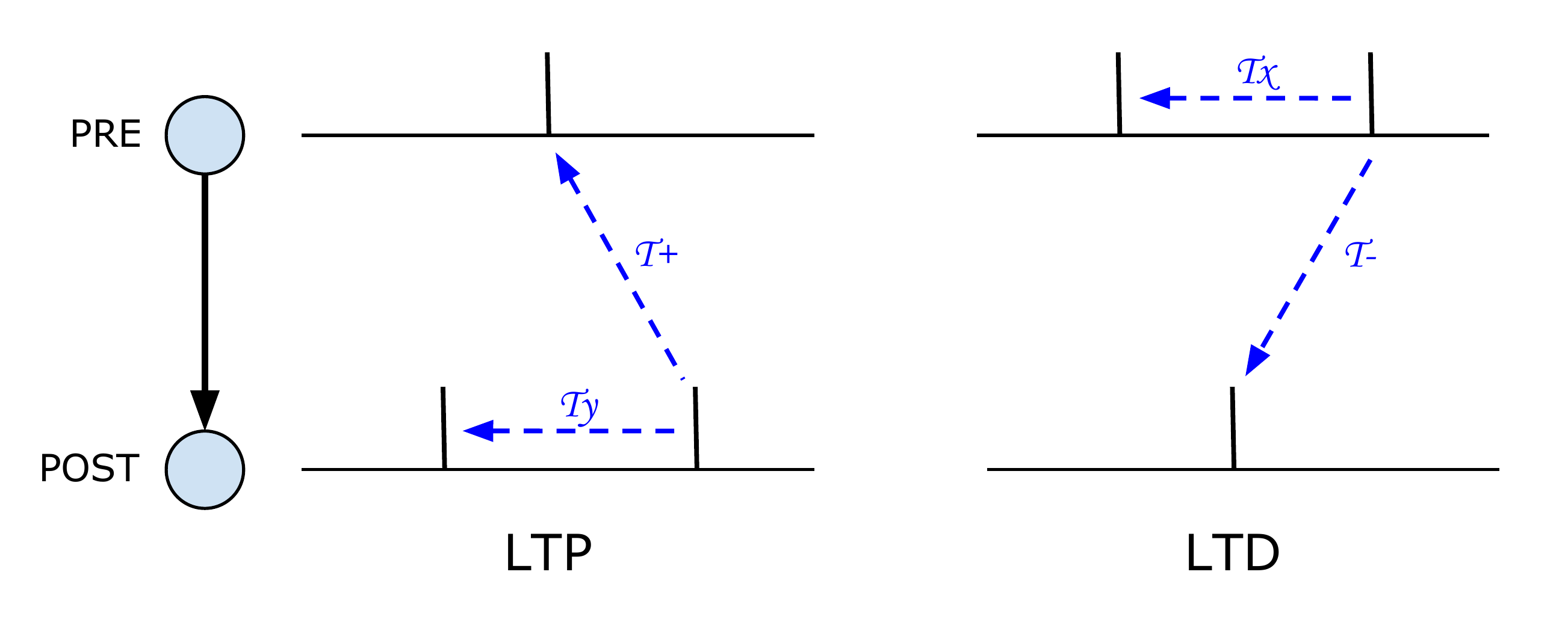}
				\caption{
				    Mechanism of tSTDP. Potentiation is achieved when the pre-synaptic unit fires in between two
				    post-synaptic spikes, whereas depression occurs when the post-synaptic unit fires in between
				    two pre-synaptic spikes.				
				}
				\label{fig10}
			\end{figure}

			A model for tSTDP is implemented in the following way.
			The mechanism is similar to that of pSTDP in the sense that the difference between spike times in pre- and
			post-synaptic neurons determines the amount of synaptic modulation, however one extra term is added to the
			weight update function.
			This extra term considers the temporal difference between the two most recent spikes of one of the spiking
			nodes. Hence, the spike triplet.
			Potentiation occurs in a similar way as in pSTDP, however the amount of synaptic modulation is also in function
			of the difference between the two most recent spikes of the post-synaptic unit. Therefore, this rule is
			identified as \emph{post-pre-post}.
			Similarly, depression occurs analogously to pSTDP, but the synaptic update is in function of the temporal difference
			between the two most recent spikes of the pre-synaptic neuron. So, the rule is summarized as \emph{pre-post-pre}.
			Figure \ref{fig10} shows a schematic representation of this two rules.
			
			Thus, in order to extend pSTDP into tSTDP we add some extra terms to the equation for 
			$\Delta w_{ij}(\Delta t)$ described in Sect.~\ref{model:STDP}, which yields:
			
			\[ \Delta w_{ij}(\Delta t) = \left\{
			\begin{array}{l l}
				a_{p} \exp\{\frac{-\Delta T_{1}}{T_{p}}\} \exp\{\frac{-\Delta T_{2}}{T_{y}}\} & \quad \text{if $\Delta T_{1}\geq 0$}\\
				-a_{d} \exp\{\frac{\Delta T_{1}}{T_{d}}\} \exp\{\frac{-\Delta T_{2}}{T_{x}}\} & \quad \text{if $\Delta T_{1}< 0$}
			\end{array} \right.\]		
			
            \noindent where parameters $a_{p}$ and $T_{p}$ set the amount and duration of LTP, whereas 
			$a_{d}$ and $T_{d}$ set the amount and duration of LTD, as with pSTDP (see above);
			$T_{1}$ represents the difference between pre- and post-synaptic spikes,
			$T_{2}$ denotes the temporal difference between the two most immediate post-synaptic spikes (if $\Delta T_{1}\geq 0$)
			or the temporal difference between the most immediate pre-synaptic spikes (if $\Delta T_{1}< 0$);
			and $T_{x}$ and $T_{y}$ are two parameters that in a similar fashion as with parameters $T_{d}$ and $T_{p}$
			set the amount of influence of immediate spikes for depression and potentiation, respectively.
			
			With these ideas in mind we report the following observations when considering tSTDP mechanisms in our model
			rather than simple pSTDP.
			Similar to pSTDP the synaptic modulation triggered by tSTDP mechanisms alter the quality of the power-law approximation to the distribution
			of avalanche sizes. As expected, this in turn has effects over the error function and the exponent of the power-law fit.
			This behaviour is also captured by the the largest eigenvalue $\Lambda$ of the weight matrix $W$.
			Similar to pSTDP, when tSTDP sets in the synaptic modulation takes $\Lambda$ away from unity
			reflecting the deviation from criticality induced by tSTDP  (see Table~\ref{tabLargstEig}).

			However, although tSTDP has an effect on the modulation of the synaptic strength between two connected nodes, this type of plasticity is less effective 
			than pSTDP on severing connections; to the point that almost all edges survive after simulation time.
			Because of this, we do not observe any small-world structure emerging from fully-connected networks as we did with pSTDP.
			At criticality, tSTDP acts as a mild modulator of synaptic weights based on the firing activity of pre- and post-synaptic nodes.

			Why tSTDP does not achieve the same amount of pruning as pSTDP?
			Let node $i$ send an edge to node $j$; and let $j$ spike before $i$.
			For potentiation to take place, a second spike from $j$ to take place.
			Let us refer to the first spike of $j$ as $j_{1}$, and to its second spike as $j_{2}$, whereas the the only spike of $i$ will be denoted as $i_{1}$.
			If node $j$'s first spike and node $i$'s only spike belong to the same avalanche we write $j_{1}, i_{1}\in Av_{1}$.
			As said above, for potentiation to occur, a second spike from $j$ must take place. These leaves the following possibilities:
			\begin{enumerate}
				\item $j_{1}, i_{1}\in Av_{1}$ and $j_{2}\in Av_{2}$. Here, for the maximum potentiation to occur the inter-avalanche interval must be very small.
				Such a regime, in which firing rate is high, is associated with super-criticality and not with criticality.
				\item $j_{1}\in Av_{1}$ and $i_{1}, j_{2}\in Av_{2}$. Same as above. Here spikes are separated in two different avalanches that require to be as close
				as possible in order to potentiation be at its fullest.
				\item $j_{1}, i_{1}, j_{2}\in Av_{1}$. Here what is required is that the three spikes occur in the same avalanche. For this to happen node $j$ must be active
				twice in a single avalanche giving place to a non-Hamiltonian avalanche, that is, an avalanche in which there is a node becomes active more than once.
				Although not shown here, this occurs naturally in scale-free and random networks at criticality due to the sparseness of the network.
				Therefore we expect this to happen in this type of networks and to be non-existent in fully-connected structures.
	 		\end{enumerate}
			
			We observe that the third case occurs in scale-free networks and this will result in weights reaching the largest allowed value of synaptic
			weight ($w_{max}$), whereas fully-connected networks fall in cases ($1$) and ($2$) described above, in which the weights do not settle in the maximum synaptic
			weight and rather give rise to a bell-shaped distribution for the weights (not shown here).
			
			Let us consider the case where $i$ fires before $j$. For depression to take place we require a second spike from node $i$.
			Similar to the case for potentiation, we have the following possibilities:
			\begin{enumerate}
				\item $i_{1}, j_{1}\in Av_{1}$ and $i_{2}\in Av_{2}$. As with the cases presented above, this situation requires a small inter-avalanche interval in order
				to get the maximum possible amount of synaptic depression.
				\item $i_{1}\in Av_{1}$ and $j_{1}, i_{2}\in Av_{2}$.  Same as above.
				\item $i_{1}, j_{1}, i_{2}\in Av_{1}$. As with potentiation, this case requires non-Hamiltonian avalanches to be present in the system. As mentioned earlier,
				only scale-free and random estructures exhibit this kind of behaviour at criticality.
			\end{enumerate}
			
			Unlike potentiation, we do not observe a large amount of synaptic depression occurring in our networks, and consequently no synaptic pruning either.
			This is due to the asymmetry of the STDP learning rule, in which potentiation is benefited over depression plus the fact that the rule requires either
			a small inter-avalanche interval in the system or non-Hamiltonian avalanches.

	\section{Discussion}
		\label{discussion}
		
		    One important observation regarding small-world-ness in networks is that of the case of
		    fully-connected topologies.
		    It can be argued that this particular type of structure possesses the small-world property 
		    \emph{ab initio} as it has both the maximum value of 
		    mean clustering coefficient and the lowest value of mean path length possibles,
		    therefore any alteration to its structure can only
		    impair its degree of small-world-ness.
		    This is a very valid observation. 
		    However, we should point out that a fully-connected network is a blank slate, in which no 
		    dynamic process has yet taken place.
		    STDP carves a structure out of this topology (in particular, pSTDP), that changes the ratio of the mean path length and the mean clustering coefficient when compared
		    to a random network with the same number of nodes and edges.
		    The resulting structure is better than the one that would have emerged
		    from a process pruning edges randomly.
		    Moreover, the value of $S$ for a fully-connected structure is $1$ always, as the
		    comparison with a random network with same number of nodes and edges yields exactly
		    the same fully-connected structure.
		    However, we observe that after STDP the fully-connected network becomes a
		    network in which clearly $S>1$ (see Fig.~\ref{fig1:a}), which implies the presence
		    of the small-world property.
		   
		    The model of Basalyga et al.~\cite{basalyga2011emergence} is,
		    unlike ours, not based on critical dynamics.
		    They consider Erd\"os-Renyi networks of slightly more complex unit but
		    only at a network size of $N=100$ neurons, while we tried to consider 
		    scaling relations for system sizes larger than that.
		    Since in there model only excitatory synapses
		    are subject to STDP, the topological effects are biased. While the main
		    observation that STDP can impair criticality is reproduced here, we find
		    a clear improvement of the small-world-ness of the emerging networks.
		
		    Our analysis includes an inspection of the motif profile that resulted from STDP mechanisms.
		    Song et al.~\cite{song2005highly} estimated the motif distribution of acute slices 
		    from the visual cortex of rats and observed that the motif profile of such networks differs 
		    from their random counterparts, in particular bidirectional connections were found to be more
		    frequent than expected by chance. 
		    In contrast, we have obtained a slight decrease of bidiretional motifs which is related to the particular form of the STDP rule used here \cite{shin2006self}
		    when compared against a random network, which is not our case. 
		    As Song et al. claim, their counts are relative to random, whereas ours are absolute counts,
		    that is, real counts which are not compared against a random network~\cite{song2005highly}.
		    In their work, they present a ratio of actual counts to that predicted by their null 
		    hypothesis.
		    In this context, bidirectional connections show an overrepresentation, a situation which is 
		    not in clash with our observations.
		    When the ratio of motifs containing bidirectional connections is compared to that of 
		    non-bidirectional connections in the same network profile, 
		    we observe that the latter is much
		    larger than the former, an observation that is in agreement with ours. 
		    If we compare the resulting state of our model to the state before
		    STDP was applied, we find that many bidirectional connections have been deleted.
		    Moreover, it can be said that from a certain point of view, their observations are just 
		    snapshots of a dynamic process happening in the networks that they
		    considered. Such a process would have another effect in the long run, in which bidirectional
		    connections are found more rarely than before.
		    Also, these authors do not mention anything related to the type of dynamic process that gave
		    rise to the structures that they observed.
		    Thus, the question regarding STDP as a mechanism that prunes two-way connections is not 
		    settled by their work.

		    \emph{Ex nihilo nihil fit.} 
		    The brain is shaped not only by genetics 
		    but also by activity-dependent processes during development. 
		    It is known that several brain structures possess the features commonly associated to 
		    small-world networks~\cite{sporns2010networks}. In particular,
		    the only neural network that has been mapped in its entirety, namely the nervous
		    system of the nematode \emph{C.~elegans}, is known to
		    possess the small-world property~\cite{watts1998collective}.
		    In higher organisms, genetic processes cannot fully account for the existence of 
		    this particular network structure, as 
		    during the lifetime of an individual, modification of the network are present due to 
		    activity-dependent processes, e.g. in connection to learning and memory. 
		    Here, we claim that massively connected structures combined with critical dynamics can
		    give rise to a small-world structure already if a standard STDP rule
		    is in place for adapting the network towards a \emph{better-than-random} structure which
		    is beneficial for information transmission across the system.
\begin{acknowledgments}
    VHU would like to thank the Mexican National Council on Science and 
	Technology (CONACYT) fellowship no. 214055 for partially funding this work.
\end{acknowledgments}

	\bibliography{biblio}

	    \begin{turnpage}
	        \begin{table}[t]
                \begin{tabular}{llrrrr}
                    \hline
                    Type & Subtype & Size & $\Lambda_{static}$ & $\Lambda_{pSTDP}$ & $\Lambda_{tSTDP}$\\
                    \hline
\multirow{8}{*}{Out-degree scale-free} \\
& Low Mean CC & $128$ & $0.906\pm 0.029$ & $0.27\pm 0.009$ & $0.35\pm 0.01$ \\
 & & $256$ & $0.9\pm 0.02$ & $0.45\pm 0.007$ & $0.52\pm 0.016$\\
 & & $512$ & $0.95\pm 0.01$ & $0.7\pm 0.005$ & $0.85\pm 0.002$\\
		\cline{2-6}
 & High Mean CC & $128$ & $0.89\pm 0.04$ & $0.26\pm 0.01$ & $0.33\pm 0.02$\\
 & & $256$ & $0.91\pm 0.03$ & $0.43\pm 0.01$ & $0.51\pm 0.01$\\
 & & $512$ & $0.91\pm 0.01$ & $0.63\pm 0.01$ & $0.73\pm 0.009$\\
                    \hline
\multirow{8}{*}{In-degree scale-free} \\ 
& Low Mean CC & $128$ & $0.98\pm 0.02$ & $0.21\pm 0.005$ & $0.32\pm 0.009$ \\
 & & $256$ & $0.99\pm 0.01$ & $0.4\pm 0.006$ & $0.49\pm 0.01$\\
 & & $512$ & $1.0006\pm 0.006$ & $0.62\pm 0.006$ & $0.68\pm 0.002$\\
		\cline{2-6}
 & High Mean CC & $128$ & $0.96\pm 0.02$ & $0.23\pm 0.013$ & $0.307\pm 0.02$\\
 & & $256$ & $0.99\pm 0.02$ & $0.408\pm 0.016$ & $0.47\pm 0.01$\\
 & & $512$ & $1.002\pm 0.014$ & $0.61\pm 0.01$ & $0.62\pm 0.007$\\
                     \hline
Random & & $128$ & $0.92\pm 0.012$ & $0.28\pm 0.004$ & $0.34\pm 0.003$ \\
 & & $256$ & $0.99\pm 0.022$ & $0.48\pm 0.003$ & $0.53\pm 0.004$\\
 & & $512$ & $0.97\pm 0.001$ & $0.77\pm 0.001$ & $0.87\pm 0.001$\\
 			\hline
Fully-connected & & $128$ & $0.91\pm 0.034$ & $0.098\pm 0.007$ & $0.63\pm 0.0001$ \\
 & & $256$ & $0.93\pm 0.0006$ & $0.23\pm 0.0062$ & $0.78\pm 0.0002$\\
 & & $512$ & $0.95\pm 0.001$ & $0.33\pm 0.004$ & $0.88\pm 0.0005$\\
                    \hline
                \end{tabular}
                \caption{
                		Largest eigenvalue $\Lambda$ of matrix $W$ of synaptic weights.
			It has been found analytically that $\Lambda = 1$ is associated with a system at criticality~\cite{larremore2011predicting}.
			The synaptic weight matrices of our networks have $\Lambda\approx 1$ before starting STDP mechanisms.
			Once STDP (either paired- or triplet-wise) mechanisms set in the synaptic weights are modulated based on neuronal activity.
			This affects the spectral nature of the weight matrix such that $\Lambda\ll 1$ post-STDP.
			Columns $\Lambda_{static}$, $\Lambda_{pSTDP}$ and $\Lambda_{tSTDP}$ refer to the largest eigenvalue $\Lambda$ for
			pre-STDP, post-paired-wise STDP and post-triplet-wise STDP, respectively.
			(We present mean values and standard deviations.)              
                }
                \label{tabLargstEig}
            \end{table}
        \end{turnpage}

\end{document}